\documentclass[10pt,letterpaper,twocolumn]{article} 

\usepackage{ol}
\usepackage{hyperref}
\usepackage{amsmath}

\begin{document}

\twocolumn[ 

\title{Squeezing at 946nm with periodically-poled KTiOPO$_4$}



\author{Takao Aoki, Go Takahashi, and Akira Furusawa}

\address{Department of Applied Physics, School of Engineering, The University of Tokyo, 7-3-1 Hongo, Bunkyo-ku, Tokyo 113-8656, Japan and CREST, Japan Science and Technology (JST) Agency, 1-9-9 Yaesu, Chuo-ku, Tokyo 103-0028, Japan}


\begin{abstract}
We report generation of squeezed vacuum in sideband modes of continuous-wave light
at 946nm using a periodically poled KTiOPO$_4$ crystal in an optical parametric oscillator. 
At the pump power of 250mW, we observe the squeezing level of $-5.6 \pm 0.1$dB and the anti-squeezing level of $+12.7 \pm 0.1$dB.
The pump power dependence of the observed squeezing/anti-squeezing levels agrees with the theoretically calculated values 
when the phase fluctuation of locking is taken into account.
\end{abstract} 


 ] 

\noindent

Suppressed quantum noise of squeezed light can improve the sensitivity of optical measurements 
that is otherwise limited by the shotnoise\cite{Polzik92a,Caves81}. In addition to such application 
to precision measurements, squeezing gives rise to the altered interaction of atoms and 
light\cite{Gardiner86,Turchette98,Rice92}. Another application of squeezed light is in the field of 
continuous-variable quantum information science. Squeezed states are utilized to generate 
continuous-variable entanglement or to perform quantum nondemolition measurements\cite{Braunstein05}. 

Parametric down conversion processes in subthresould optical parametric oscillators (OPOs) is 
often employed to generate continuous-wave squeezed light.
Squeezing over $-6$dB has been observed by operating OPOs with nonlinear crystals in non-critical 
phase matching condition (e.g. KNbO$_3$\cite{Polzik92b}, LiNbO$_3$\cite{Schneider98}). 
Although non-critical phase matching enables efficient nonlinear optical couplings, it is possible only 
in a limited wavelength region. On the other hand, quasi phase matching in periodically-poled 
materials\cite{Fejer92} allows efficient nonlinear optical couplings in a broad range of wavelength. 
Therefore, these materials are utilized for wavelength conversion\cite{Myers95} and also for generation of squeezed 
light\cite{Anderson95,Serkland95,Serkland97,Hirano05}.

The stability of lasers is essential for stable generation of highly squeezed light.
In this regard, Diode-pumped Nd:YAG lasers at 1064nm have excellent stability and they have been widely 
used for generation of squeezed light\cite{Schneider98,Lam99,McKenzie04}. 
In these experiments, 
InGaAs photodiodes are commonly used as light-detecting devices. The quantum efficiency of these 
devices achievable with current technology is about 95\%, which results in a 5\% detection loss. 
In contrast, Si photodiodes with quantum efficiencies over 99\% are available in the wavelength region 
from visible to about 950nm. 
Hence the use of Nd:YAG laser at 946nm\cite{Fan87} and Si photodiodes will reduce the detection loss 
in these systems. Here we report generation of squeezed vacuum at 946nm using periodically-poled 
KTiOPO$_4$ (PPKTP).

\begin{figure}[htb]
\centerline{\includegraphics[width=8cm]{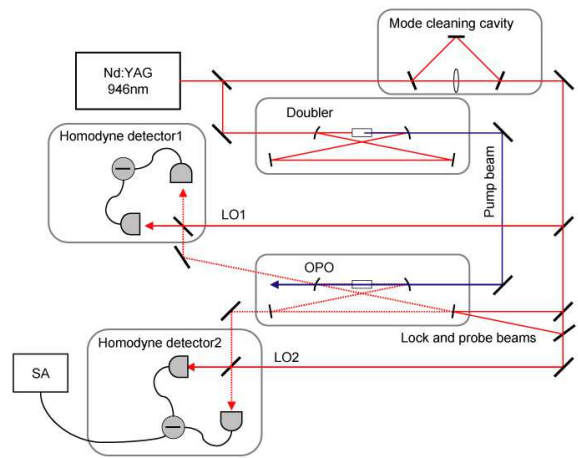}}
\caption{Experimental setup. LO: local oscillator, OPO: optical parametric oscillator, SA: spectrum analyzer.
Homodyne detector1 is used to measure the amplitude of the lock beam to lock the OPO cavity length. 
AC output of the homodyne detector2 is put into a spectrum analyzer to measure the squeezing/anti-squeezing 
levels, while DC output is used to lock the LO phase and pump beam phase relative to the probe beam.
}
\end{figure}

Figure 1 shows the experimental setup. 
We use a continuous-wave diode-pumped monolithic Nd:YAG laser at 946nm (Innolight Mephisto QTL) with an
output power of 500 mW. The second harmonic of the laser is generated in an external-cavity frequency 
doubler. The frequency doubler has a bow-tie type ring configuration with two spherical mirrors 
(radius of curvature of 25mm) and two flat mirrors. One of the spherical mirrors has a reflectivity 
of 90\% at 946nm and is used as an input coupler, while the others are high-reflectivity coated. 
All the mirrors have reflectivities of less than 5\% at 473nm.
A 10mm-long PPKTP crystal (Raicol Crystals) is used as a nonlinear crystal for second harmonics generation. 
The cavity length is actively stabilized using the tilt-locking method\cite{Shaddock99}.
The input 946nm beam is slightly misaligned (a few percent in power) in the horizontal direction 
to have a sufficient error signal.
The generated 473nm beam pumps a degenerate optical parametric oscillator (OPO). 
The OPO also has a bow-tie type ring configuration with two spherical mirrors (radius of curvature of 25mm) 
and two flat mirrors. One of the flat mirrors has reflectivity of 85\% at 946nm and is used as an output coupler. 
The round-trip cavity length is 214 mm, which results in a waist radius of 17$\mu$m inside the crystal.  
Again, a 10mm-long PPKTP crystal is used as a nonlinear crystal for parametric down conversion. 
The OPO is driven below the parametric oscillation threshold to generate squeezed vacuum states.
20mW of 946nm beam from the Nd:YAG laser is spatially filtered in a mode-cleaning cavity with a 
triangle type ring configuration. The output beam (8mW) is split into four beams: a probe beam, a locking beam, 
and two local oscillator beams for homodyne detection.
The probe beam is injected into the OPO cavity through a high-reflection flat mirror. The transmitted 
probe beam from the output coupler (100$\sim$500nW at parametric gain of 1) is detected with a balanced-homodyne detector.
The balanced-homodyne detector has two Si photodiodes (HAMAMATSU S3590-06, anti-reflective coated at 946nm) with 
quantum efficiency of 99.4\% at 946nm. The photocurrents are directly subtracted and put into the AC branch 
electronics to measure squeezing and the DC branch electronics to lock the local oscillator phase and to measure 
and lock the parametric gain.  
The locking beam is also injected into the cavity through a high-reflection flat mirror in the mode
counter-propagating to the probe beam. The amplitude of the transmitted beam (10$\mu$W) 
is measured with a homodyne detector.
The error signal for the dither-locking of the cavity length is extracted from the measured amplitude.
The whole setup is on a 750mm $\times$ 1200mm breadboard (Newport VH3048W-OPT-28).

Figure 2 shows the measured noise levels at the pump power of 250mW as the local oscillator phase 
is (i) scanned, (ii) locked at the anti-squeezed quadrature, and (iii) locked at the squeezed quadrature 
compared to the shotnoise level (iv). The noise level is measured with a spectrum analyzer in the zero 
span mode at 1MHz, with the resolution bandwidth of 30kHz and the video bandwidth of 300Hz. The traces 
are averaged for 30 measurements except for (i). The squeezing level of $-5.6\pm 0.1$dB and the anti-squeezing level 
of $+12.7\pm 0.1$dB are observed. By subtracting the detector circuit noise, we obtain the inferred 
squeezing/anti-squeezing levels of $-5.80\pm 0.1$dB and $+12.72\pm 0.1$dB, respectively.

The variance of the output mode $R_{\pm}$ for the anti-squeezed ($+$) and the squeezed ($-$) quadratures 
can be modeled as\cite{Polzik92b,Collett84}
\begin{align}
R_{\pm} &= 1 \pm \alpha \rho \frac{4x}{(1 \mp x)^2 + 4\Omega^2}, 
\end{align}
where $\alpha$ and $\rho$ are the detection efficiency and the OPO escape efficiency, respectively. 
The detection efficiency $\alpha$ is a product of the propagation efficiency $\zeta$, 
the photodiode quantum efficiency $\eta$, and the homodyne efficiency $\xi^2$ ($\xi$ is the 
visibility between the output and the local oscillator modes), 
$\alpha = \zeta \eta \xi^2$. The OPO escape efficiency can be written as
\begin{align}
\rho = \frac{T}{T+L},
\end{align}
where $T$ and $L$ are the transmission of the output coupler and the intracavity loss, respectively.
The pump parameter $x$ is related to the classical parametric amplification gain $G$ as
\begin{align}
G &= \frac{1}{(1 - x)^2}.
\end{align}
The detuning parameter $\Omega$ is given as the ratio of the measurement frequency $\omega$ to the OPO 
cavity decay rate $\gamma = c(T+L)/l$ ($l=$ the cavity round trip length),
\begin{align}
\Omega &= \frac{\omega}{\gamma}.
\end{align}

\begin{figure}[htb]
\centerline{\includegraphics[width=8cm]{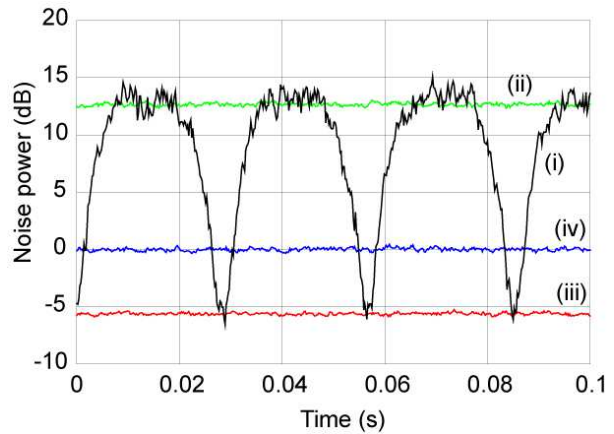}}
\caption{Measured noise levels at the pump power of 250mW. (i) LO phase is scanned. 
(ii) LO phase is locked at anti-squeezed quadrature. (iii) LO phase is locked at squeezed quadrature.
(iv) Shotnoise level. Noise levels are displayed as the relative power compared to the shotnoise level (0dB).
The settings of the spectrum analyzer are, zero-span mode at 1MHz, resolution bandwidth = 30kHz, video 
bandwidth = 300Hz. All the traces except (i) are averaged for 30 times. Detector circuit noise is not corrected.}
\end{figure}

In the current setup, $\zeta \approx 1$, $\eta = 0.994$, $\xi = 0.979$, therefore $\alpha = 0.953$. 
$T = 0.15$ and $L = 0.011$ yield $\rho = 0.932$. It should be noted that KTP and PPKTP crystals often suffer 
from the absorption induced by the pump light\cite{Wang04} as is the case with other nonlinear crystals 
(e.g., KNbO$_3$\cite{Mabuchi94}). However, our crystal makes no measurable change of the intracavity loss in the 
presence of the pump light. We measure the classical parametric amplification gain of $G=8.83$. The detuning 
parameter is $\Omega=0.028$. With these values, eq. (1) predicts the theoretical squeezing/anti-squeezing levels 
of $-8.20$dB and $+13.27$dB. Although the theoretical value for the anti-squeezing is consistent with the experiment, 
there is a discrepancy between them for the squeezing. 

This discrepancy may be explained by taking into account the phase fluctuation of the locking\cite{Zhang03}.
Assuming that the relative phase between the local oscillator and the anti-squeezed/squeezed quadratures has 
a normal distribution with a small standard deviation of $\tilde \theta$, the noise levels to be observed can 
be written as
\begin{align}
R^{\prime}_{\pm}(\tilde \theta) &= \int\frac{1}{\sqrt{2\pi}\tilde \theta} %
                                   \exp \left( - \frac{\theta ^2}{2\tilde \theta ^2}\right) %
                                   \left( R_{\pm} \cos^2 \theta \, + R_{\mp} \sin^2 \theta \right) d\theta
                                   \nonumber \\
                                &\approx R_{\pm} \cos^2 {\tilde \theta} + R_{\mp} \sin^2 {\tilde \theta} .
\end{align}
Therefore, phase fluctuation with an rms of $\tilde \theta$ is effectively equivalent to having a phase 
offset of $\tilde \theta$.
 
From the measurement on the rms noise of the error signal of locking circuits, we obtain the total rms phase 
fluctuation of $\tilde \theta_{total} = 4.3 \pm 0.6^{\circ}$. This results in the corrected theoretical 
noise levels of $-5.68\pm 0.56$dB and $+13.25\pm 0.1$dB, which are in good agreement with the experimentally 
observed values. 

\begin{figure}[htb]
\centerline{\includegraphics[width=8cm]{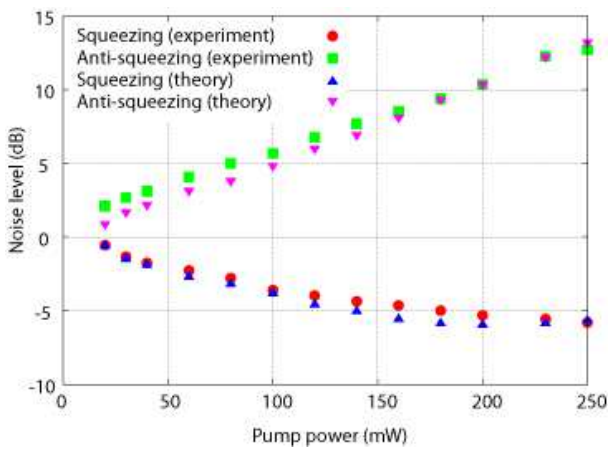}}
\caption{Pump power dependence of the observed squeezing/anti-squeezing levels (circles/squares) and calculated 
values (upper/lower triangles) according to eq. (5). Detector circuit noise is corrected for experimentally 
observed levels. }
\end{figure}

We repeat the above measurement and analysis for various pump powers. The results are summarized in Fig. 3.
Theoretical values fairly agree with the experimental results. 
The reason for the slight difference in the values of anti-squeezing at low pump powers is not fully understood.
Furthermore, we perform the same experiments as Fig. 3 on another PPKTP crystal.
We again observe no absorption change induced by the pump light and obtain similar measurement results 
($-5.73\pm 0.1$dB/$+12.22\pm 0.1$dB for squeezing/anti-squeezing levels after correction of the detector circuit noise).

In summary, we observe $-5.6\pm 0.1$dB squeezing and $+12.7 \pm 0.1$dB anti-squeezing with PPKTP in an OPO.
The pump power dependence of the observed squeezing/anti-squeezing levels agree with theoretical model with 
the phase fluctuation of the locking. Reducing the phase fluctuation by stabilizing the setup (both 
actively and passively) may help to observe better squeezing levels.

This work was partly supported by the MPHPT and the MEXT of Japan.



\begin{thebibliography}{99}
\bibitem{Polzik92a} E.S. Polzik, J. Carri, and H.J. Kimble, Phys. Rev. Lett {\bf 68,} 3020 (1992).

\bibitem{Caves81} C.M. Caves, Phys. Rev. D {\bf 23,} 1693 (1981).

\bibitem{Gardiner86} C.W. Gardiner, Phys. Rev. Lett {\bf 56,} 1917 (1986).

\bibitem{Turchette98} Q.A. Turchette, N.Ph. Georgiades, C.J. Hood, H.J. Kimble, and A.S. Parkins,
Phys. Rev. A {\bf 58,} 4056 (1998).

\bibitem{Rice92} P.R. Rice and L.M. Pedrotti, J. Opt. Soc. Am. B {\bf 9,} 2008 (1992).

\bibitem{Braunstein05} S.L. Braunstein and P. van Loock, Rev. Mod. Phys. {\bf 77,} 513 (2005).

\bibitem{Polzik92b} E.S. Polzik, J. Carri, and H.J. Kimble, Appl. Phys. B {\bf 55,} 279 (1992).

\bibitem{Schneider98} K. Schneider, M. Lang, J. Mlynek, and S. Schiller, Optics Express {\bf 2,} 59 (1998).

\bibitem{Fejer92} M.M. Fejer, G.A. Magel, D.H. Jundt, and R.L. Byer, IEEE J. Quant. Elctron. {\bf 28,} 2631 (1992).

\bibitem{Myers95} L.E. Myers, R.C. Eckardt, M.M. Fejer, R.L. Byer, W.R. Bosenberg, and J.W. Pierce,
J. Opt. Soc. Am. B {\bf 12,} 2102 (1995).

\bibitem{Anderson95} M.E. Anderson, M. Beck, M.G. Raymer, and J.D. Bierlein, Opt. Lett. {\bf 20,} 620 (1995).

\bibitem{Serkland95} D.K. Serkland, M.M. Fejer, R.L. Byer, and Y. Yamamoto, Opt. Lett. {\bf 20,} 1649 (1995).

\bibitem{Serkland97} D.K. Serkland, P. Kumar, M.A. Arbore, and M.M. Fejer, Opt. Lett. {\bf 22,} 1497 (1997).

\bibitem{Hirano05} T. Hirano, K. Kotani, T. Ishibashi, S. Okude, and T. Kuwamoto, Opt. Lett. {\bf 30,} 1722 (2005).

\bibitem{Lam99} P.K. Lam, T.C. Ralph, B.C. Buchler, D.E. McClelland, H.A. Bachor, and J. Gao, J. Opt. B {\bf 1,} 469 (1999). 

\bibitem{McKenzie04} K. McKenzie, N. Grosse, W.P. Bowen, S.E. Whitcomb, M.B. Gray, D.E. McClelland, and P.K. Lam,
Phys. Rev. Lett. {\bf 93,} 161105 (2004).

\bibitem{Fan87} T.Y. Fan and R.L. Byer, Opt. Lett. {\bf 12,} 809 (1987).

\bibitem{Shaddock99} D.A. Shaddock, M.B. Gray, and D.E. McClelland, Opt. Lett. {\bf 24,} 1499 (1999).

\bibitem{Collett84} M.J. Collett and C.W. Gardiner, Phys. Rev. A {\bf 30,} 1386 (1984).

\bibitem{Wang04} S. Wang, V. Pasiskevicius, and F. Laurell, J. Appl. Phys. {\bf 96,} 2023 (2004).

\bibitem{Mabuchi94} H. Mabuchi, E.S. Polzik, and H.J. Kimble, J. Opt. Soc. Am. B {\bf 11,} 2023 (1994).

\bibitem{Zhang03} T.C. Zhang, K.W. Goh, C.W. Chou, P. Lodahl, and H.J. Kimble, Phys. Rev. A {\bf 67,} 033802 (2003).

\end{thebibliography}
\end{document}